\documentclass[pra,aps,twocolumn]{revtex4}

\usepackage{bm}
\usepackage{epsfig}
\usepackage{amssymb}

\begin{document}

\title{Minimal disturbance measurement for coherent states is non-Gaussian}
\author{Ladislav Mi\v{s}ta Jr.}
\affiliation{Department of Optics, Palack\' y University, 17.
listopadu 50,  772~07 Olomouc, Czech Republic}
\date{\today}

\begin{abstract}
In standard coherent state teleportation with shared two-mode squeezed 
vacuum (TMSV) state there is a trade-off between the teleportation 
fidelity and the fidelity of estimation of the teleported state 
from results of the Bell measurement. Within the class of Gaussian operations 
this trade-off is optimal, i.e. there is not a Gaussian operation which would 
give for a given output fidelity a larger estimation fidelity. We 
show that this trade-off can be improved by up to $2.77\%$ if we use a 
suitable non-Gaussian operation. This operation can be implemented by 
the standard teleportation protocol in which the shared TMSV state 
is replaced with a suitable non-Gaussian entangled state. We also 
demonstrate that this operation can be used to enhance the transmission 
fidelity of a certain noisy channel.
\end{abstract}

\maketitle

\section{Introduction}
In quantum mechanics there is not an operation which would give 
some information on an unknown quantum state without disturbing 
the state. This property of quantum mechanics is closely related 
to the no-cloning theorem \cite{Wootters_82} which forbids to 
perfectly duplicate an unknown quantum state. A question that can 
be risen in this context is which operation allowed by quantum mechanics 
approximates best this non-existing operation, i.e. which operation introduces
for a given information gain the least possible disturbance. 
Naturally, this operation, conventionally denoted as minimal 
disturbance measurement (MDM), will in general depend on the set of input 
states, their a-priori distribution, and on the figures of 
merit used to quantify the information gain and the state disturbance. 
A convenient approach to the problem on finding the MDM was developed 
in \cite{Banaszek_01a}. In this approach the classical information 
gained on the input state from a quantum operation is converted into 
the estimate of the input state and the information gain is then 
quantified by the average estimation fidelity $\bar{G}$, i.e. the fidelity 
$G$ between the estimate and the input state averaged over the distribution 
of the input states. On the other hand, the disturbance introduced 
by the operation into the input state is quantified by the 
average output fidelity $\bar{F}$, i.e. the fidelity $F$ between the input 
state and the state at the output of the operation averaged over the 
distribution of the input states. According to the laws of quantum
mechanics for a given set of input states there exists a specific optimal
trade-off between the fidelities $\bar{G}$ and $\bar{F}$ which cannot be overcome
by any quantum operation. In terms of the fidelities $\bar{G}$ and
$\bar{F}$ the MDM then can be defined as a quantum operation which saturates
this optimal trade-off. First optimal fidelity trade-offs and the corresponding 
MDMs were derived in the context of finite-dimensional quantum 
systems and observables with discrete spectra. To be more specific, 
the MDMs were found analytically for a completely unknown \cite{Banaszek_01a} 
and partially known \cite{Mista_05} $d$-level particle and numerically for $N$ 
identical copies of a completely unknown $2$-level particle (qubit) \cite{Banaszek_01b}. 
In addition, the MDMs for a completely unknown as well as partially 
known qubit were demonstrated experimentally using a single-photon polarization 
qubit \cite{Sciarrino_05}. Besides being of fundamental interest MDM can be applied 
to increase the transmission fidelity of a certain lossy channel \cite{Ricci_05}.

Only recently the concept of MDM was also extended into the realm of systems with 
infinite-dimensional Hilbert spaces and observables with continuous spectrum-continuous 
variables (CVs). The attention has been paid to MDMs on Gaussian states, i.e. states 
represented by Gaussian Wigner function, realized by covariant Gaussian operations, 
i.e. operations preserving Gaussian states which are invariant under displacement 
transformations. These operations are advantageous since for coherent input states 
they posses state-independent output fidelity $F$ and estimation fidelity $G$ which 
can be conveniently used for characterization of state disturbance and information gain. 
Within the class of such operations optimal trade-off between the two fidelities 
as well as the corresponding MDM for the set of all coherent states with uniform 
a-priori distribution were derived in \cite{Andersen_05,Fiurasek_05} and realized 
experimentally in \cite{Andersen_05}. In addition, also this covariant Gaussian MDM 
was shown to be capable to increase transmission fidelity of some noisy channels 
\cite{Andersen_05}.

In this paper we address a natural question of whether the covariant Gaussian 
MDM for uniformly distributed coherent states can be improved. 
We answer this question in the affirmative. We show that the fidelity 
trade-off corresponding to this measurement can be increased by up to $2.77\%$ 
if we use a suitable non-Gaussian operation thus showing that MDM for 
coherent states is non-Gaussian. This MDM can be implemented using the 
standard continuous-variable (CV) teleportation protocol \cite{Vaidman_94,Braunstein_98}
in which the participants share an appropriate non-Gaussian entangled state. 
Further, we demonstrate that our non-Gaussian MDM gives a higher transmission 
fidelity of a certain noisy channel in comparison with that achieved in 
\cite{Andersen_05}. As a by-product we also derive for the set of all 
coherent states with uniform a-priori distribution a lower bound for the 
optimal fidelity trade-off for any covariant quantum operation. The present 
paper is inspired by the recent result that fidelity of quantum cloning of 
coherent states can be increased by using a non-Gaussian entangled 
state \cite{Cerf_04}.

The paper is organized as follows. Section~\ref{sec_1} deals with optimal 
Gaussian fidelity trade-off and corresponding MDM for uniformly distributed 
coherent states. In Section~\ref{sec_2} we derive for this set of states a 
lower bound on optimal fidelity trade-off for any covariant quantum operation. 
Section~\ref{sec_3} is dedicated to implementation of a quantum operation 
saturating this bound and its application. Section~\ref{sec_4} contains conclusions.
\section{Gaussian minimal disturbance measurement}\label{sec_1}

The Gaussian MDM can be realized by at least three ways \cite{Andersen_05} encompassing 
the asymmetric cloning followed by a joint measurement, a linear-optical scheme with  
a feed forward or by the standard CV teleportation protocol proposed by Braunstein and 
Kimble (BK) \cite{Braunstein_98} and demonstrated experimentally in \cite{Furusawa_98}. 
With respect to what follows it is convenient to start with the implementation of the 
Gaussian MDM via BK teleportation protocol. Here we use an optical notation in which 
CV systems are realized by single modes of an optical field and the role of CVs is 
played by quadratures $x_{j}$ and $p_{k}$ ($[x_{j},p_{k}]=i\delta_{jk}$) of these modes. 

\begin{figure}
\centerline{\psfig{width=8.0cm,angle=0,file=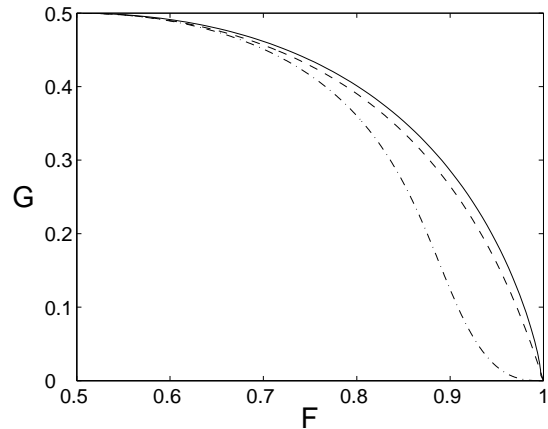}}
\caption{Trade-off between the otput fidelity $F$ and the
estimation fidelity $G$ for the BK teleportation scheme with the
shared optimized non-Gaussian state (solid curve), TMSV state (\ref{TMSV}) (dashed curve) and 
TMSV state de-gaussificated by local single photon subtraction from each mode (\ref{subtracted}) 
(dotted-dashed curve). See text for details.}
\label{fig1}
\end{figure}

In the BK protocol an unknown coherent state $|\alpha\rangle_{\rm in}$
of a mode ``in'' is teleported by sender Alice ($A$) to receiver Bob ($B$). 
In each run of the protocol, the state is chosen randomly with uniform 
distribution from the set of all coherent states. Initially, Alice 
and Bob share a two-mode squeezed vacuum (TMSV) state which is a Gaussian entangled 
state of two optical modes $A$ and $B$ described in the Fock basis as
\begin{equation}\label{TMSV}
|TMSV\rangle_{AB}=\sum_{n=0}^{\infty}\tilde c_{n}|n,n\rangle_{AB},
\quad \tilde c_{n}=\sqrt{1-\lambda^2}\lambda^{n}, 
\end{equation}
where $\lambda=\tanh r$ ($r$ is the squeezing parameter). Then, Alice performs 
the so called Bell measurement that consists of superimposing of modes ``in'' and $A$ 
on a balanced beam splitter and subsequent detection of the quadrature variables 
$x_{1}=(x_{\rm in}-x_{A})/\sqrt{2}$ and $p_{2}=(p_{\rm in}+p_{A})/\sqrt{2}$ 
at its outputs. She obtains classical results of the measurement $\bar x_{1}$ and $\bar p_{2}$ 
and sends them via classical channel to Bob who displaces his part of the shared state as 
$x_{B}\rightarrow x_{\rm out}=x_{B}+\sqrt{2}\bar x_{1}$ and 
$p_{B}\rightarrow p_{\rm out}=p_{B}+\sqrt{2}\bar p_{2}$. As a result, Bob's output quadratures 
read as $x_{\rm out}=x_{\rm in}-\sqrt{2}e^{-r}x_{B}^{(0)}$ and 
$p_{\rm out}=p_{\rm in}+\sqrt{2}e^{-r}p_{A}^{(0)}$ \cite{Loock_00a},
where $x_{i}^{(0)}$ and $p_{j}^{(0)}$ stand for initial vacuum
quadratures. Since the squeezing $r$ is always finite in practice Bob has only an approximate 
replica $\rho_{\rm out}(\alpha)$ of the input state. Moreover, for
the same reason Alice gains some information on the input state from results of the Bell measurement 
that can be converted into a classical estimate $\rho_{\rm est}(\alpha)$ of the input state by displacing 
a vacuum mode $E$ as $x_{E}^{(0)}\rightarrow x_{\rm est}=x_{E}^{(0)}+\sqrt{2}\bar x_{\rm 1}$ and 
$p_{E}^{(0)}\rightarrow p_{\rm est}=p_{E}^{(0)}+\sqrt{2}\bar p_{\rm 2}$. Hence 
she obtains $x_{\rm est}=x_{\rm in}+x_{E}^{(0)}-(e^{r}x_{A}^{(0)}+e^{-r}x_{B}^{(0)})/\sqrt{2}$ 
and $p_{\rm est}=p_{\rm in}+p_{E}^{(0)}+(e^{-r}p_{A}^{(0)}+e^{r}p_{B}^{(0)})/\sqrt{2}$.
Quantifying now the resemblance of the states $\rho_{\rm out}(\alpha)$ and $\rho_{\rm est}(\alpha)$ 
to the input state $|\alpha\rangle$ by the output fidelity $F$ and the estimation fidelity $G$
\begin{equation}\label{fidelities}
F=\langle\alpha|\rho_{\rm out}(\alpha)|\alpha\rangle,\quad 
G=\langle\alpha|\rho_{\rm est}(\alpha)|\alpha\rangle, 
\end{equation}
one finds using the latter formulas that  
\begin{equation}\label{FGBK}
F_{\rm BK}=\frac{1}{1+e^{-2r}},\quad G_{\rm BK}=\frac{1}{1+\cosh^{2}(r)}.
\end{equation}
Expressing $\cosh^{2}(r)$ in terms of $F_{\rm BK}$ using the first formula and inserting this into 
the formula for $G_{\rm BK}$ we finally arrive at the following trade-off 
between the fidelities $F_{\rm BK}$ and $G_{\rm BK}$:  
\begin{equation}\label{trade-off}
G_{\rm BK}=\frac{1}{1+\frac{1}{4F_{\rm BK}(1-F_{\rm BK})}}.
\end{equation}
The obtained trade-off is depicted by dashed curve in Fig.~\ref{fig1}. It is obvious 
from the figure that as one would expect the obtained fidelities exhibit 
complementary behavior, i.e. the larger is the estimation fidelity the smaller is 
the output fidelity and vice versa. Interestingly, it was shown in 
\cite{Andersen_05,Fiurasek_05} that if one restricts only to the covariant 
Gaussian operations, then the trade-off (\ref{trade-off}) is optimal. It means 
in other words that the standard BK teleportation protocol with shared TMSV 
state realizes (within the class of all covariant Gaussian operations) MDM 
for coherent states. In the following sections we demonstrate that the optimal Gaussian trade-off 
(\ref{trade-off}) can be improved by a suitable covariant non-Gaussian operation. 
\section{Optimal fidelity trade-off for coherent states}\label{sec_2}

We start by a suitable mathematical formulation of the task on finding the optimal 
fidelity trade-off for CVs. For this purpose we use a general method developed in 
\cite{Mista_05}. We restrict our attention to 
coherent input states $|\alpha\rangle=D(\alpha)|0\rangle$, where $\alpha$ lies in 
the complex plane $\mathbb{C}$, which form the orbit of the Weyl-Heisenberg group. 
Here $|0\rangle$ is the vacuum state and the displacement operators 
$D(\alpha)=\mbox{exp}(\alpha a^{\dag}-\alpha^{\ast}a)$, $\alpha\in\mathbb{C}$, where  
$a$ and $a^{\dag}$ are the standard annihilation and creation
operators satisfying boson commutation rule $[a,a^{\dag}]=\openone$, comprise the irreducible 
unitary representation of this group. We also assume that the a-priori distribution of the input 
states coincides with the invariant measure on the group 
$d^{2}\alpha/\pi=d(\mbox{Re}\alpha)d(\mbox{Im}\alpha)/\pi$.

The standard BK teleportation protocol can be formally viewed as 
a trace-preserving quantum operation which is covariant, 
i.e. invariant under displacement transformations, and which can give as a 
measurement outcome any complex number $\beta$ 
($\beta\equiv\bar x_{1}+i\bar p_{2}$ in the case of
teleportation). Therefore, we will seek for the optimal fidelity trade-off 
on this set of quantum operations. To each outcome $\beta$ of such an 
operation we can assign a trace-decreasing completely positive (CP) 
map that can be represented by the following positive-semidefinite operator on 
the tensor product ${\mathcal H}_{\rm in}\otimes{\mathcal H}_{\rm out}$ of the 
input and output Hilbert spaces $\mathcal{H}_{\rm in}$ and $\mathcal{H}_{\rm out}$ 
\cite{Jamiolkowski_72}
\begin{equation}\label{covariant}
\chi(\beta)=[D_{\rm in}(\beta^{\ast}) \otimes D_{\rm out}(\beta)] \chi_{0}  
[D_{\rm in}^{\dag}(\beta^{\ast}) \otimes D_{\rm out}^{\dag}(\beta)],
\end{equation}
where $\chi_{0}$ is a positive-semidefinite operator. For the measurement outcome 
$\beta$ the estimated state is $|\beta\rangle$ whereas the output state reads
\begin{equation}\label{out}
\rho(\beta|\alpha)=\mathrm{Tr}_{\rm in}[\chi(\beta) 
(|\alpha\rangle_{\rm in}\langle\alpha|)^{\rm T}\otimes \openone_{\rm out}],
\label{rhoout}
\end{equation}
where $\openone_{\rm out}$ is the identity operator on $\mathcal{H}_{\rm out}$. Since the map 
$\chi(\beta)$ is trace-decreasing the output state (\ref{rhoout}) is not normalized to unity 
and its norm $P(\beta|\alpha)=\mathrm{Tr}_{\rm out}[\rho(\beta|\alpha)]$ is equal to 
the probability density of this outcome. The entire operation should be trace-preserving 
which imposes the following constraint:
\begin{equation}\label{constraint}
\frac{1}{\pi}\int_{\mathbb{C}}\mathrm{Tr}_{\rm out}[\chi(\beta)]d^{2}\beta =\openone_{\rm in},
\end{equation}
where $\int_{\mathbb{C}}$ denotes integration over the whole complex plane 
and $\openone_{\rm in}$ is the identity operator on $\mathcal{H}_{\rm in}$.
For the input state $|\alpha\rangle$ the studied operation produces 
on average the output state and the estimated state in the form
\begin{eqnarray}\label{rho}
\rho_{\rm out}(\alpha)&=&\frac{1}{\pi}\int_{\mathbb{C}}\rho(\beta|\alpha)d^{2}\beta,\nonumber\\
\rho_{\rm est}(\alpha)&=&\frac{1}{\pi}\int_{\mathbb{C}} P(\beta|\alpha)|\beta\rangle\langle\beta|d^{2}\beta.
\end{eqnarray}
Making use of Eqs.~(\ref{covariant}), (\ref{out}), (\ref{rho}) and the formula
$D(-\beta)|\alpha\rangle=\mbox{exp}[i\mbox{Im}(\alpha\beta^{\ast})]|\alpha-\beta\rangle$
one finally finds the fidelities (\ref{fidelities}) to be
\begin{equation}
F= \mathrm{Tr}[\chi_{0}R_F], \qquad G= \mathrm{Tr}[\chi_{0}R_G],
\label{FGchi}
\end{equation}
where $R_{F}$ and $R_{G}$ are the positive-semidefinite Gaussian operators defined as
\begin{eqnarray}\label{R}
R_F&=&\frac{1}{\pi}\int_{\mathbb{C}}|\gamma^{\ast}\rangle_{\rm in}\langle\gamma^{\ast}|\otimes
|\gamma\rangle_{\rm out}\langle\gamma|d^{2}\gamma, \nonumber\\
R_G&=&\frac{1}{\pi}\int_{\mathbb{C}}e^{-|\gamma|^{2}}|\gamma\rangle_{\rm in}\langle\gamma|
d^{2}\gamma\otimes\openone_{\rm out}.  
\end{eqnarray}
The optimal trade-off between the fidelities $F$ and $G$ can be found by finding the 
maximum of the weighted sum 
\begin{equation}\label{transmission}
\mathcal{F}(p)=pF+(1-p)G
\end{equation}
of these two fidelities 
\cite{Fiurasek_03}, where the parameter $p\in[0,1]$ controls the ratio between the 
information gained from the input state and the disturbance of this state. We can write 
$\mathcal{F}(p)=\mathrm{Tr}[\chi_0 R(p)]$, where
\begin{equation}\label{Rp}
R(p)=p R_F+(1-p)R_G.
\end{equation}
Making use of the inequality $R(p)\leq\lambda_{\rm max}(p)(\openone_{\rm in}\otimes\openone_{\rm out})$,
where $\lambda_{\rm max}(p)$ is the maximum eigenvalue of $R(p)$ and taking into account the condition
$\mathrm{Tr}[\chi_{0}]=1$ which we obtain from the constraint (\ref{constraint}) 
using the formula \cite{D'Ariano_04} 
\begin{equation}\label{Schur}
\frac{1}{\pi}\int_{\mathbb{C}}D(\alpha)XD^{\dag}(\alpha)d^{2}\alpha=\mathrm{Tr}[X]\openone 
\end{equation}
following from Schur's lemma, one finds $\mathcal F(p)$ to be upper bounded as 
$\mathcal{F}(p)\leq\lambda_{\rm max}(p)$. Now, if we find a normalized eigenvector
$|\chi_{\rm max}(p)\rangle$ of the operator $R(p)$ corresponding to the maximum 
eigenvalue $\lambda_{\rm max}(p)$, then the map (\ref{covariant}) generated from 
any such $\chi_{0,{\rm max}}=|\chi_{\rm max}(p)\rangle\langle\chi_{\rm max}(p)|$ 
satisfies the trace-preservation condition (\ref{constraint}) as follows 
from Eq.~(\ref{Schur}). Consequently, the map $\chi_{0,{\rm max}}$ is the 
optimal one that saturates optimal trade-off between $F$ and $G$.

The finding of the optimal fidelity trade-off thus boils down to the diagonalization 
of the operator $R(p)$ which acts on the direct product of two infinite-dimensional 
spaces. For this purpose it is convenient to express the operators $R_F$ and $R_G$ 
in the form:
\begin{eqnarray}\label{explicitR}
R_F&=&\sum_{K=0}^{\infty}\frac{{K!}}{2^{K+1}}\sum_{n,m=0}^{K}\frac{|n\rangle_{\rm in}\langle K-m|
\otimes|m\rangle_{\rm out}\langle K-n|}{\sqrt{{n!}{(K-m)!}{m!}{(K-n)!}}}, \nonumber \\
R_G&=&\sum_{n=0}^{\infty}\frac{1}{2^{n+1}}|n\rangle_{\rm in}\langle
n|\otimes\openone_{\rm out},  
\end{eqnarray}
which can be calculated from Eq.~(\ref{R}) using the formulas 
$|\gamma\rangle=e^{-|\gamma|^{2}/2}\sum_{n=0}^{\infty}(\gamma^{n}/\sqrt{n!})|n\rangle$
and $\int_{\mathbb{C}}e^{-s|\gamma|^{2}}\gamma^{n}{\gamma^{\ast}}^{m}d^{2}\gamma=
(\pi{n!}/s^{n+1})\delta_{nm}$ (Re$s>0$). The present eigenvalue problem can be simplified, 
if we notice that both the operators (\ref{explicitR}) and therefore also the operator 
$R(p)$ commute with the operator of the photon number difference 
$N_{-}=n_{\rm in}-n_{\rm out}$, where $n_{i}=a_{i}^{\dag}a_{i}$, $i={\rm in},{\rm out}$. 
Consequently, the total Hilbert space splits into the direct sum 
$\displaystyle {\mathcal H}_{\rm in}\otimes{\mathcal H}_{\rm out}=
\mathop{\oplus}_{N=-\infty}^{+\infty}{\mathcal H}^{(N)}$ of the characteristic 
subspaces ${\mathcal H}^{(N)}$ of the operator $N_{-}$ corresponding to the eigenvalues 
$N=-\infty,\ldots,+\infty$. The infinite-dimensional subspaces 
${\mathcal H}^{(+L)}$ (${\mathcal H}^{(-L)}$), $L=0,1,\ldots$ are spanned by the basis vectors 
$\{|n+L,n\rangle_{\rm in,out}, n=0,1,\ldots\}$ ($\{|n,n+L\rangle_{\rm in,out}, n=0,1,\ldots\}$).
Hence, it remains to diagonalize the operator $R(p)$ in the subspaces ${\mathcal H}^{(\pm L)}$ 
where it is represented by infinite-dimensional matrices 
$R^{(\pm L)}(p)=pR_{F}^{(\pm L)}+(1-p)R_{G}^{(\pm L)}$, where 
\begin{eqnarray}\label{RL}
\left(R_F^{(\pm L)}\right)_{nm}&=&\frac{\sqrt{\left({n+m+L \atop n}\right)\left({n+m+L \atop m}\right)}}
{2^{L+n+m+1}}, \nonumber \\
\left(R_G^{(+L)}\right)_{nm}&=&\frac{\delta_{nm}}{2^{n+L+1}},\quad \left(R_G^{(-L)}\right)_{nm}=
\frac{\delta_{nm}}{2^{n+1}},\nonumber\\ 
\end{eqnarray}
where $n,m=0,1,\ldots$. 

\begin{figure}
\centerline{\psfig{width=8.0cm,angle=0,file=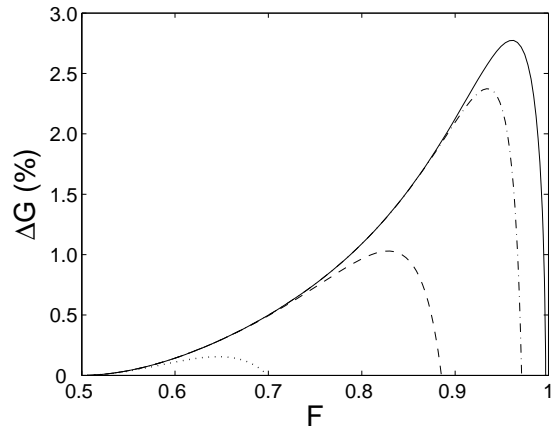}}
\caption{Difference $\Delta G=G-G_{\rm BK}$ between the estimation fidelity 
$G$ in the improved fidelity trade-off and the estimation fidelity 
$G_{\rm BK}$ in the optimal Gaussian trade-off versus the output fidelity $F$ for $\mathcal{N}=4$ (dotted curve), 
$\mathcal{N}=12$ (dashed curve), $\mathcal{N}=50$ (dotted-dashed curve), $\mathcal{N}=500$ (solid curve). 
See text for details.}
\label{fig2}
\end{figure}

We have accomplished this task numerically and the obtained fidelity trade-off is depicted by the solid 
curve in Fig.~\ref{fig1}. The figure clearly demonstrates that this trade-off beats the optimal 
Gaussian trade-off (\ref{trade-off}) (dashed curve). In order to see the degree of improvement better 
we have plotted by a solid curve in Fig.~\ref{fig2} the dependence of difference $\Delta G=G-G_{\rm BK}$ 
between the estimation fidelity $G$ in the improved trade-off and the estimation fidelity 
$G_{\rm BK}$ in the optimal Gaussian trade-off on the output fidelity $F$. Numerical analysis reveals that, 
for instance, $\Delta G\approx 1.04\%$ is achieved for $F\approx 0.794$ and the maximum improvement 
of $\Delta G\approx 2.77\%$ is attained for $F=F_{\rm max}\approx 0.963$. 

It should be stressed that we have in fact found a lower bound on the optimal trade-off because 
we have approximated each original infinite-dimensional matrix $R^{(\pm L)}(p)$ by its 
$\mathcal{N}$-dimensional submatrix $R_{\mathcal{N}}^{(+L)}(p)$ ($R_{\mathcal{N}}^{(-L)}(p)$) 
on the $\mathcal{N}$-dimensional subspace ${\mathcal H}_{\mathcal{N}}^{(+L)}$ 
(${\mathcal H}_{\mathcal{N}}^{(-L)}$) spanned by the basis vectors 
$\{|n+L,n\rangle_{\rm in,out}, n=0,1,\ldots,\mathcal{N}-1\}$ 
($\{|n,n+L\rangle_{\rm in,out}, n=0,1,\ldots,\mathcal{N}-1\}$).
This follows from the inequality  
\begin{equation}\label{inequality}
\langle\psi_{\mathcal{N}}|R_{\mathcal{N}}^{(\pm L)}(p)|\psi_{\mathcal{N}}\rangle
=\langle\psi_{\mathcal{N}}|R(p)|\psi_{\mathcal{N}}\rangle\leq\lambda_{\rm max}(p), 
\end{equation}
which holds for any $|\psi_{\mathcal{N}}\rangle\in{\mathcal H}_{\mathcal{N}}^{(\pm L)}$. 
Therefore, because we only calculated the eigenvector corresponding to maximum eigenvalue 
of matrices $R_{\mathcal{N}}^{(\pm L)}(p)$ (here we took $\mathcal{N}=500$ and $L\leq 30$), 
the optimal trade-off can be slightly larger than that given by the solid curve in Fig.~\ref{fig1}.     
\section{Implementation}\label{sec_3}

The above analysis shows that the optimal eigenvector $|\chi_{\rm max}(p)\rangle$ 
lies in one of the subspaces ${\mathcal H}^{(\pm L)}$. We have a strong
numerical evidence that it lies in the subspace corresponding to $L=0$, i.e. it 
has the structure 
\begin{equation}\label{chimax}
|\chi_{\rm max}(p)\rangle=\sum_{n=0}^{\infty}c_{n}|n,n\rangle_{AB},
\end{equation}
and, in addition, the probability amplitudes $c_{n}$ comprising the dominant 
eigenvector of the matrix $R^{(0)}(p)$ are nonnegative. The latter
statement follows immediately from positivity of elements of the matrix $R(p)$. The
optimal CP map (\ref{covariant}) corresponding to the vector (\ref{chimax}) can be 
implemented by the BK teleportation scheme in which the TMSV state (\ref{TMSV}) is replaced 
by this vector. This can be shown if we describe the BK teleportation by the transfer operator 
method \cite{Hofmann_01}. In this formalism, the action of the BK teleporter 
with shared entangled state $\sum_{n=0}^{\infty}c_{n}|n,n\rangle_{AB}$ is described by the set of 
transfer operators $\{T(\beta)=D(\beta)T(0)D^{\dag}(\beta)\}_{\beta\in\mathbb{C}}$,
where $T(0)=\sum_{n=0}^{\infty}c_{n}|n\rangle\langle n|$. If the input state is $|\alpha\rangle$ 
and the Bell measurement gives the outcome $\beta=\bar x_{1}+i\bar p_{2}$
the output state reads as $|\psi_{\rm out}(\beta|\alpha)\rangle=T(\beta)|\alpha\rangle$.
The state $|\chi_{0}\rangle$ forming the positive-semidefinite operator $\chi_{0}$ 
describing the considered teleportation protocol then can be calculated by acting 
with the operator $T(0)$ on one part of maximally entangled state 
$\sum_{n=0}^{\infty}|n,n\rangle_{AB}$ \cite{Jamiolkowski_72}. This finally gives 
the state $|\chi_{0}\rangle=T_{B}(0)\sum_{n=0}^{\infty}|n,n\rangle_{AB}$, 
which coincides exactly with the state (\ref{chimax}). 

\begin{figure}
\centerline{\psfig{width=8.0cm,angle=0,file=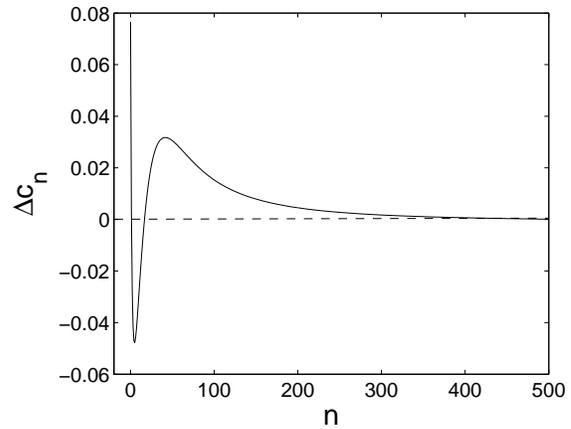}}
\caption{Dependence of the difference $\Delta c_{n}=c_{n}-\tilde c_{n}$ of the Schmidt
coefficient $c_{n}$ of the entangled state (\ref{suboptimal}) with $\mathcal{N}=500$ 
and the Schmidt coefficient $\tilde c_{n}$ of the TMSV state (\ref{TMSV}) for 
$F=F_{\rm max}\approx 0.963$ on the photon number.}
\label{fig3}
\end{figure}

Likewise, we can implement the quantum operation which saturates the fidelity 
trade-off depicted by the solid curve in Fig.~\ref{fig1}. For this purpose, we need 
to prepare the entangled state 
\begin{equation}\label{suboptimal}
\sum_{n=0}^{\mathcal{N}-1}c_{n}|n,n\rangle_{AB},
\end{equation}
where nonnegative probability amplitudes $c_{n}$ form the dominant eigenvector of the matrix 
$R_{\mathcal{N}}^{(0)}(p)$. Apparently, the improvement $\Delta G$ which can be achieved 
when using the state (\ref{suboptimal}) will vary with the dimension $\mathcal{N}$ of the 
truncated space ${\mathcal H}_{\mathcal{N}}^{(0)}$. This dependence is depicted in Fig.~\ref{fig2}. 
We see from the figure that the maximum improvement increases and moves towards larger values 
of $F$ as $\mathcal{N}$ grows. It is also seen from the figure that in order to achieve 
$\Delta G\approx 1\%$ one needs at least $\mathcal{N}\geq 12$. For values of $\mathcal{N}$ 
where the improvement $\Delta G$ achieves at least a few tenths of percent one can calculate 
the probability amplitudes $c_{n}$ of the state (\ref{suboptimal}) only numerically. In order 
to demonstrate the difference between the state (\ref{suboptimal}) and the optimal 
Gaussian state (\ref{TMSV}) we display in Fig.~{\ref{fig3}} the difference 
$\Delta c_{n}=c_{n}-\tilde c_{n}$ of the Schmidt coefficient $c_{n}$ of the state 
(\ref{suboptimal}) with $\mathcal{N}=500$ and the Schmidt coefficient $\tilde c_{n}$ 
of the TMSV state (\ref{TMSV}) for $F=F_{\rm max}\approx 0.963$ versus the photon number. 
The state (\ref{suboptimal}) can be prepared, at least in principle, using the probabilistic 
scheme for preparation of an arbitrary two-mode state with finite Fock state expansion based on 
linear optics \cite{Kok_02}.

The specific feature of the optimal state (\ref{chimax}) is that it possesses 
perfect correlations in photon number as the TMSV state (\ref{TMSV}). 
However, there is a sharp difference between the two states, because in contrast with 
the latter state the former one is non-Gaussian.
To show this assume on the contrary that the state (\ref{chimax}) is a Gaussian 
state of two modes $A$ and $B$. Such a state is completely characterized by the first moments 
$\langle\xi_k\rangle=\langle\chi_{\rm max}(p)|\xi_k|\chi_{\rm max}(p)\rangle$, where 
$\xi=(x_{A},p_{A},x_{B},p_{B})^{\rm T}$, and by the variance matrix $V$ with elements 
$V_{kl}=\langle\{\Delta \xi_{k},\Delta\xi_{l}\}\rangle$, 
where $\Delta\xi_{k}=\xi_{k}-\langle\xi_{k}\rangle$ and $\{A,B\}\equiv(1/2)(AB+BA)$.
As $\langle\xi\rangle=0$ for the state (\ref{chimax}), it is completely described  
just by the variance matrix which reads as  
\begin{eqnarray}\label{standard}
V=\left(\begin{array}{cccc}
a & 0 & c & 0 \\
0 & a & 0 & -c \\
c & 0 & a & 0 \\
0 & -c & 0 & a\\
\end{array}\right),
\end{eqnarray}
where $a=\sum_{n=0}^{\infty}nc_{n}^{2}+1/2$ and $c=\sum_{n=0}^{\infty}(n+1)c_{n}c_{n+1}\geq0$.
Taking into account the purity of the state, which imposes the constraint 
$\sqrt{\mbox{det}V}=a^{2}-c^{2}=1/4$ \cite{Trifonov_97}
we see that such a state would be a TMSV state (\ref{TMSV}) with 
$\lambda=\sqrt{(a-1/2)/(a+1/2)}$ which does not beat the trade-off of the 
BK scheme and thus we arrive to a contradiction. Therefore, the state (\ref{chimax}) 
is inevitably non-Gaussian. Thus we have found a non-Gaussian operation which possesses 
a better trade-off between output and estimation fidelities than any covariant Gaussian 
operation which implies that MDM for a completely unknown coherent state is non-Gaussian.

It can be interesting to compare the fidelity trade-off derived by us with the trade-off that 
would be obtained when teleporting with the state produced by local 
single photon subtraction from each mode of a TMSV state. Originally investigated in the 
context of increase of teleportation fidelity via local operations and classical 
communication \cite{Opatrny_00,Cochrane_02,Olivares_03} the state was also shown to be 
suitable for loophole-free Bell test based on homodyne detection \cite{Nha_04,Garcia-Patron_04,Olivares_04,Garcia-Patron_05}.  
The reason for studying the state here is twofold. First, the state is a non-Gaussian state of the form 
\cite{Cochrane_02,Garcia-Patron_05} 
\begin{equation}\label{subtracted}
\sqrt{\frac{(1-T^{2}\lambda^{2})^{3}}{1+T^{2}\lambda^{2}}}\sum_{n=0}^{\infty}(n+1)(T\lambda)^{n}|n,n\rangle_{AB},
\end{equation}
where $T$ is a transmittance of an unbalanced beam splitter used for photon subtraction 
and therefore it possesses perfect correlations in photon number as the optimal state (\ref{chimax}).
Second, the subtraction of a single photon was already demonstrated experimentally for a 
single-mode squeezed vacuum state \cite{Wenger_04}. Making use of the formulas 
\begin{eqnarray*}
F=\sum_{m,n=0}^{\infty}{m+n \choose n}\frac{c_m^{\ast}c_{n}}{2^{m+n+1}},\quad G=\sum_{n=0}^{\infty}\frac{|c_{n}|^{2}}{2^{n+1}} 
\end{eqnarray*}
for the teleportation and estimation fidelities in the BK teleportation with the shared 
state (\ref{chimax}) we can find using Eq.~(\ref{subtracted}) the teleportation fidelity 
to be \cite{Olivares_03}  
\begin{eqnarray*}\label{Fsubtracted}
F_{\rm s}=\frac{(1+T\lambda)^{3}(2-2T\lambda+T^{2}\lambda^{2})}{4(1+T^{2}\lambda^{2})},
\end{eqnarray*}
while the estimation fidelity reads as
\begin{eqnarray*}\label{Fsubtracted}
G_{\rm s}=2\left(\frac{2+T^{2}\lambda^{2}}{1+T^{2}\lambda^{2}}\right)\left(\frac{1-T^{2}\lambda^{2}}{2-T^{2}\lambda^{2}}\right)^{3}.
\end{eqnarray*}
The trade-off between the fidelities $F_{\rm s}$ and $G_{\rm s}$ is depicted by the dotted-dashed 
curve in Fig.~\ref{fig1}. The figure clearly reveals that the trade-off is even worse than the 
optimal Gaussian trade-off (\ref{trade-off}).  Thus while the single photon 
subtraction can be a useful method for distillation of the CV entanglement and 
test of Bell inequalities it is not suitable for preparation of a non-Gaussian 
entangled state which would improve fidelity trade-off in teleportation of coherent states.
\begin{figure}
\centerline{\psfig{width=8.0cm,angle=0,file=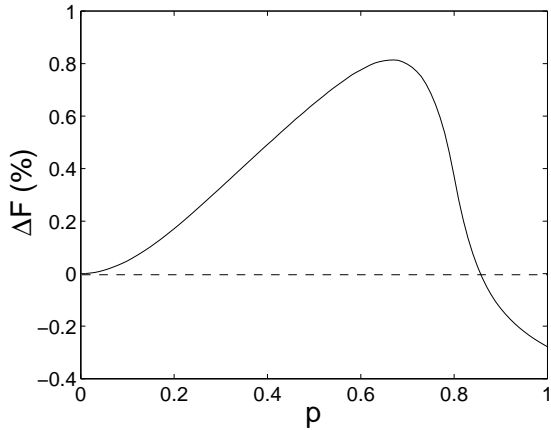}}
\caption{Dependence of the difference $\Delta F=\mathcal{F}-\mathcal{F}_{\rm BK}$ of 
the transmission fidelity $\mathcal{F}=pF+(1-p)G$ for the non-Gaussian operation 
and the transmission fidelity $\mathcal{F}_{\rm BK}=pF_{\rm BK}+(1-p)G_{\rm BK}$ 
maximized with respect to the squeezing parameter $r$ on the transmission 
probability $p$. See text for details.}
\label{fig4}
\end{figure}

The non-Gaussian operation realized by the BK teleportation protocol with shared 
non-Gaussian state (\ref{suboptimal}) can be applied to enhance 
the transmission fidelity of a certain noisy channel. The channel in question transmits perfectly 
with probability $p$ the input coherent state while with probability $1-p$ the 
state is completely absorbed by the channel. For the set of all coherent states 
with uniform a-priori distribution the channel possesses the average transmission 
fidelity equal to $\mathcal{F}_{\rm av}(p)=p$. In \cite{Andersen_05} it was demonstrated 
that for $0<p<4/5$ the transmission fidelity can be improved by using the Gaussian MDM 
in front of the channel while for $p\geq4/5$ it is better to entirely use the channel. 
Using the BK teleportation protocol to realize the MDM the improved scheme works as 
follows. Instead of sending directly the input coherent state through the channel 
one sends through it one part of the TMSV state (\ref{TMSV}). 
In the next step the other part of thus obtained state is used for teleportation of the input 
coherent state. The transmission fidelity for this scheme is given by the formula 
$\mathcal{F}_{\rm BK}(p)=pF_{\rm BK}+(1-p)G_{\rm BK}$, where the fidelities 
$F_{\rm BK}$ and $G_{\rm BK}$ are given in Eq.~(\ref{FGBK}). By maximizing the 
transmission fidelity $\mathcal{F}_{\rm BK}(p)$ with respect to the squeezing parameter 
$r$ we can reach optimal performance of the scheme when in the interval $0<p<4/5$ 
$\mathcal{F}_{\rm BK}(p)>\mathcal{F}_{\rm av}(p)$ \cite{Andersen_05}. Interestingly, 
the transmission fidelity of the channel can be further improved provided that we 
use in the BK teleportation the non-Gaussian entangled state (\ref{suboptimal}) 
($\mathcal{N}=500$) as a quantum channel. This scheme must be inevitably optimal 
since within the class of all covariant operations it is designed in such a 
way that it maximizes the quantity (\ref{transmission}) which is in fact the 
transmission fidelity of the considered channel. The dependence of the 
improvement $\Delta F(p)=\mathcal{F}(p)-\mathcal{F}_{\rm BK}(p)$ on the probability 
$p$ for our scheme is depicted in Fig.~\ref{fig4}. The figure reveals 
that for $0<p\lesssim0.85$ the scheme really allows to slightly improve the 
transmission fidelity the maximum improvement of $\Delta F\approx 0.81\%$ 
being achieved for $p=0.67$. In the region of $p\gtrsim0.85$ $\Delta F$ attains 
negative values which is a numerical artefact caused by the truncation of the 
infinite-dimensional matrix $R^{(0)}(p)$ to the finite-dimensional matrix 
$R_{\mathcal{N}}^{(0)}(p)$. Therefore, in order to achieve $\Delta F>0$ 
also for some $p\gtrsim0.85$ we would need to use in teleportation the 
state (\ref{suboptimal}) with $\mathcal{N}>500$. If this is not 
possible then for $p\gtrsim0.85$ it is better to send directly 
the input coherent state through the channel rather than to use our 
non-Gaussian operation. Thus we have illustrated also practical 
utility of the studied non-Gaussian operation for increase of the 
transmission fidelity of a specific quantum channel.
\section{Conclusions}\label{sec_4}

In conclusion, we have shown that there exists a covariant non-Gaussian 
quantum operation which gives for a completely unknown coherent state 
a better trade-off between the output fidelity and the estimation 
fidelity than any covariant Gaussian operation. This means that 
the covariant MDM for a completely unknown coherent state is 
non-Gaussian. The non-Gaussian operation can be implemented by the 
standard BK teleportation protocol with a suitable non-Gaussian 
entangled state as a quantum channel and can be utilized to enhance 
the transmission fidelity of a certain channel. As a by-product we 
also derived a lower bound for the optimal fidelity 
trade-off for a completely unknown coherent state within the class 
all covariant quantum operations. Our result thus clearly illustrates 
that one can extract more information on an unknown coherent state 
while preserving the degree of disturbance introduced into it by 
this procedure by using a suitable non-Gaussian operation. 
\acknowledgments

I would like to thank Jarom\'{\i}r Fiur\'a\v{s}ek, Radim Filip, Ulrik Andersen, 
Miroslav Gavenda, and Radek \v{C}elechovsk\'y for valuable
discussions. The research has been supported by the research project: 
``Measurement and Information in Optics,'' No. MSM 6198959213 and by the 
COVAQIAL (FP6-511004) and SECOQC (IST-2002-506813) projects of the sixth 
framework program of EU.


\begin{thebibliography}{99}

\bibitem{Wootters_82} W. K. Wootters and W. H. Zurek, Nature (London)
{\bf 299}, 802 (1982).
\bibitem{Banaszek_01a} K. Banaszek, Phys. Rev. Lett. {\bf 86}, 1366 (2001).
\bibitem{Mista_05} L. Mi\v{s}ta, Jr., J. Fiur\'a\v{s}ek, and R. Filip,
Phys. Rev. A {\bf 72}, 012311 (2005).
\bibitem{Banaszek_01b} K. Banaszek and I. Devetak, Phys. Rev. A {\bf 64}, 052307 (2001).
\bibitem{Sciarrino_05} F. Sciarrino, M. Ricci, F. De Martini, R. Filip, and L. Mi\v{s}ta, Jr., 
e-print quant-ph/0510097.  
\bibitem{Ricci_05} M. Ricci, F. Sciarrino, N. J. Cerf, R. Filip, J. Fiur\'{a}\v{s}ek, and F. De Martini, 
Phys. Rev. Lett. {\bf 95}, 090504 (2005).
\bibitem{Andersen_05} U. L. Andersen, M. Sabuncu, R. Filip, and G. Leuchs, e-print quant-ph/0510195. 
\bibitem{Fiurasek_05} J. Fiur\'a\v{s}ek (private communication).
\bibitem{Vaidman_94} L. Vaidman, Phys. Rev. A {\bf 49}, 1473 (1994).
\bibitem{Braunstein_98} S. L. Braunstein and H.  J. Kimble, Phys. Rev. Lett. {\bf 80}, 869 (1998).
\bibitem{Cerf_04} N. J. Cerf, O. Kr\"{u}ger, P. Navez, R. F. Werner, and M. M. Wolf, 
Phys. Rev. Lett. {\bf 95}, 070501 (2005).
\bibitem{Furusawa_98} A. Furusawa, J. L. S{\o}rensen, S. L. Braunstein, C. A. Fuchs, 
H. J. Kimble, and E. S. Polzik, Science {\bf 282}, 706 (1998); W. P. Bowen, N. Treps, B. C. Buchler, 
R. Schnabel, T. C. Ralph, H.-A. Bachor, T. Symul, and P. K. Lam, Phys. Rev. A {\bf 67}, 032302 (2003);
T. C. Zhang, K. W. Goh, C. W. Chou, P. Lodahl, and H. J. Kimble, Phys. Rev. A {\bf 67}, 033802 (2003).
\bibitem{Loock_00a} P. van Loock and S. L. Braunstein, Phys. Rev. Lett. {\bf 84}, 3482 (2000).
\bibitem{Jamiolkowski_72} A. Jamiolkowski, Rep. Math. Phys. {\bf 3}, 275 (1972); M.-D. Choi, Linear 
Algebr. Appl. {\bf 10}, 285 (1975).
\bibitem{Fiurasek_03} J. Fiur\'a\v{s}ek, Phys. Rev. A {\bf 67}, 052314 (2003);
J. Fiur\'a\v{s}ek, R. Filip, and N. J. Cerf, Quant. Inf. Comp. {\bf 5}, 583 (2005). 
\bibitem{D'Ariano_04} G. M. D'Ariano, P. Perinotti, and M. F. Sacchi, J. Opt. B: Quantum 
Semiclassical Opt. {\bf 6}, 487 (2004).
\bibitem{Hofmann_01} H. Hofmann, T. Ide, T. Kobayashi, and A. Furusawa, Phys. Rev. A {\bf 64},
040301(R) (2001).
\bibitem{Kok_02} P. Kok, H. Lee, and J. P. Dowling, Phys. Rev. A {\bf
65}, 052104 (2002); J. Fiur\'a\v{s}ek, S. Massar, and N. J. Cerf,
Phys. Rev. A {\bf 68}, 042325 (2003).
\bibitem{Trifonov_97} D. A. Trifonov, J. Phys. A: Math. Gen. {\bf 30},
5941 (1997).
\bibitem{Opatrny_00} T. Opatrn\'y, G. Kurizki, and D.-G. Welsch, Phys. Rev. A {\bf 61}, 032302 (2000).
\bibitem{Cochrane_02} P. T. Cochrane, T. C. Ralph, and G. J. Milburn, Phys. Rev. A {\bf 65}, 062306 (2002).
\bibitem{Olivares_03} S. Olivares, M. G. A. Paris, and R. Bonifacio, Phys. Rev. A {\bf 67}, 032314 (2003).
\bibitem{Nha_04} H. Nha and H. J. Carmichael, Phys. Rev. Lett. {\bf 93}, 020401 (2004).
\bibitem{Garcia-Patron_04} R. Garc\'{\i}a-Patr\'on, J. Fiur\'a\v{s}ek, N. J. Cerf, J. Wenger, 
R. Tualle-Brouri, and P. Grangier, Phys. Rev. Lett. {\bf 93}, 130409 (2004).
\bibitem{Olivares_04} S. Olivares and M. G. A. Paris, Phys. Rev. A {\bf 70}, 032112 (2004).
\bibitem{Garcia-Patron_05} R. Garc\'{\i}a-Patr\'on, J. Fiur\'a\v{s}ek, and N. J. Cerf, Phys. Rev. A {\bf 71}, 
022105 (2005).
\bibitem{Wenger_04} J. Wenger, R. Tualle-Brouri, and P. Grangier, Phys. Rev. Lett. {\bf 92}, 153601 (2004).
\end{thebibliography}
\end{document}